\documentclass[aps,prl,reprint,superscriptaddress,nofootinbib,floatfix]{revtex4-2}

\usepackage[T1]{fontenc}
\usepackage{lmodern}
\usepackage{amsmath,amssymb}
\usepackage{graphicx}
\usepackage{xcolor}
\usepackage{hyperref}
\hypersetup{hidelinks}

\makeatletter
\renewcommand{\paragraph}{%
  \@startsection{paragraph}{4}{\parindent}{\z@}{-1em}%
    {\normalfont\normalsize\bfseries}%
}
\makeatother

\newcommand{\A}{\mathcal{A}}
\newcommand{\T}{\mathcal{T}}
\newcommand{\E}{\mathbb{E}_{\bx,\by}}
\newcommand{\bx}{\mathbf{x}}
\newcommand{\by}{\mathbf{y}}
\newcommand{\ket}[1]{|#1\rangle}
\newcommand{\bra}[1]{\langle #1|}
\newcommand{\braket}[2]{\langle #1|#2\rangle}

\newif\ifletterchanges
\letterchangestrue
\DeclareRobustCommand{\rev}[1]{\ifletterchanges{\color{black}#1}\else#1\fi}
\begin{document}

\title{Strong Simulation of 1D Quantum Circuits via Reduced Transition Matrices}

\author{Matilde Grassi}
\affiliation{Laboratoire de Physique Th\'eorique et Mod\'elisation, CNRS UMR 8089,
CY Cergy Paris Universit\'e, 95302 Cergy-Pontoise Cedex, France}
\affiliation{JEIP, UAR 3573 CNRS, Coll\`ege de France, PSL Research University,
11 Place Marcelin Berthelot, 75321 Paris Cedex 05, France}
\author{Stefano Carignano}
\affiliation{Barcelona Supercomputing Center, 08034 Barcelona, Spain}
\author{Luca Tagliacozzo}
\affiliation{Institute of Fundamental Physics IFF-CSIC, Calle Serrano 113b,
28006 Madrid, Spain}
\affiliation{Quantum Advanced Research Center (QuARC), CSIC, Calle Serrano 113b,
28006 Madrid, Spain}
\author{Jacopo De Nardis}
\affiliation{Laboratoire de Physique Th\'eorique et Mod\'elisation, CNRS UMR 8089,
CY Cergy Paris Universit\'e, 95302 Cergy-Pontoise Cedex, France}
\affiliation{JEIP, UAR 3573 CNRS, Coll\`ege de France, PSL Research University,
11 Place Marcelin Berthelot, 75321 Paris Cedex 05, France}
\date{\today}

\begin{abstract}
Tensor networks are powerful tools for simulating quantum many-body systems,
but the growth of spatial entanglement severely limits the direct evolution
of pure states under chaotic unitary circuits.
Here we address the task of evaluating specified output
probabilities of one-dimensional circuits at finite relative precision.
\rev{We introduce a sweeping algorithm that contracts the circuit transversely and compresses
its two temporal boundaries jointly through reduced transition matrices (RTMs),}
preserving their overlap rather than either boundary separately.
At fixed target precision, the \rev{retained} bond dimension exhibits a slowdown
consistent with subexponential growth over the accessible time window,
in a regime where the circuit is already deeply anticoncentrated.
A noisy random-operator model illustrates
how finite precision can induce a crossover from linear to logarithmic
\rev{temporal entropy growth}. These results open a route to probability queries for
benchmarking and learning without representing the full quantum state.
\end{abstract}
\maketitle

\paragraph{Introduction.}

The simulation of nonequilibrium quantum many-body states is limited by the
rapid growth of entanglement. Tensor-network methods based on matrix product
states (MPS), including time-evolving block decimation (TEBD),
the time-dependent density matrix renormalization group (tDMRG), and the
time-dependent variational principle (TDVP), mitigate this difficulty by
restricting the dynamics to low-complexity variational manifolds. These
approaches build on density matrix renormalization
\cite{white_noack1992,white1992,white1993}, adaptive real-time evolution
\cite{vidal2004,white_feiguin2004,daley2004,vidal2007,haegeman2011}, and
subsequent developments \cite{hauke2013spread,paeckel2019,banuls2023review}.
Tensor-network methods have also proved powerful tools for benchmarking
quantum processors, even outperforming current devices in accuracy and, in
some cases, speed on specific many-body simulation tasks
\cite{begusic2024fast,tindall2024eagle,ouyang2026fast,tindall2026disordered}. Their
computational cost is controlled by the spatial entanglement of the evolving
state, and they therefore eventually become inefficient when this
entanglement grows extensively \cite{vidal2003,schollwock2011}.
In many applications, however, one does not
need to represent the full time-evolved state. Here we consider the more
targeted problem of evaluating the probability of a specified measurement
outcome,
\begin{equation}
  p(\bx|\by)=|\A_{\bx\by}(T)|^2,\qquad
  \A_{\bx\by}(T)=\bra{\bx}U(T)\ket{\by},
  \label{eq:probability}
\end{equation}
to fixed \textit{relative} precision $\epsilon_p$ (or, alternatively,
to fixed relative precision $\epsilon_{\mathcal{A}}$ on the amplitude
$\mathcal{A}$).

This probability-query task is distinct from sampling the circuit's output
distribution. Sampling from sufficiently deep chaotic circuits is conjectured
to be classically hard and forms the basis of quantum-advantage experiments
\cite{aaronson2017,boixo2018,arute2019}. In suitable local random-circuit
ensembles, anticoncentration occurs at depth of order $\log N$, with typical
probabilities of order $2^{-N}$ for a system of $N$ qubits \cite{dalzell2022}.
A fixed additive error is therefore insufficient to resolve these
probabilities. Estimating them to fixed relative precision through repeated
measurements also requires exponentially many samples.

An amplitude $\A_{\bx\by}(T)$ is itself a space-time tensor network. Besides generic
contraction strategies \cite{markov2008,gray2021a,gray2024,ma2024}, one can
contract it along space, representing the left and right partial contractions
as temporal MPS (tMPS)
\cite{banuls2009,muller-hermes2012,hastings2015,frias-perez2022,lerose2021,sonner2021influence}.
For local dynamics, influence-matrix studies have found area-law temporal
entanglement in free models and logarithmic growth in interacting integrable
circuits \cite{lerose2021scaling,giudice2022temporal}; light-cone algorithms
avoid intermediate entanglement barriers \cite{lerose2023barrier}.
Exact local dynamics are also accessible in special circuit families
\cite{bertini2019correlations,pickering2026influence}.

Taken separately, the temporal boundaries can nevertheless exhibit extensive
entanglement in chaotic circuits \cite{foligno2023,ippoliti2022}.
Temporal coarse graining can reduce influence-matrix entanglement while
preserving local dynamics \cite{vilkoviskiy2025temporal}.
For a specified global amplitude, the key distinction is that only the
overlap of the boundaries is needed: approximating each faithfully is
unnecessary. This motivates reduced transition matrices (RTMs)
\cite{hastings2015,carignano2024rtm,carignano2025sampled,parzygnat2023,carignano2025cft,li2026rtm}. Here we instead address global amplitude queries at finite relative precision.

Building on these ideas, and in particular on the transverse-contraction framework of
Ref.~\cite{carignano2025itransverse}, we introduce a sweeping RTM algorithm
with internal relative-error control. \rev{Our main result is numerical evidence
\textit{that requiring a finite target precision within a RTM-based algorithm substantially reduces the resources needed for
specified amplitude queries}.} 
Specifically, we find that, for chaotic one-dimensional circuits, the accepted bond dimensions grow more
slowly than exponentially over the accessible window. We test this
behavior against exact amplitudes and Porter--Thomas statistics, and
introduce a noisy random-operator model that offers a phenomenological
explanation for the slowdown.

\begin{figure*}[t]
 \centering
 \includegraphics[width=\textwidth]{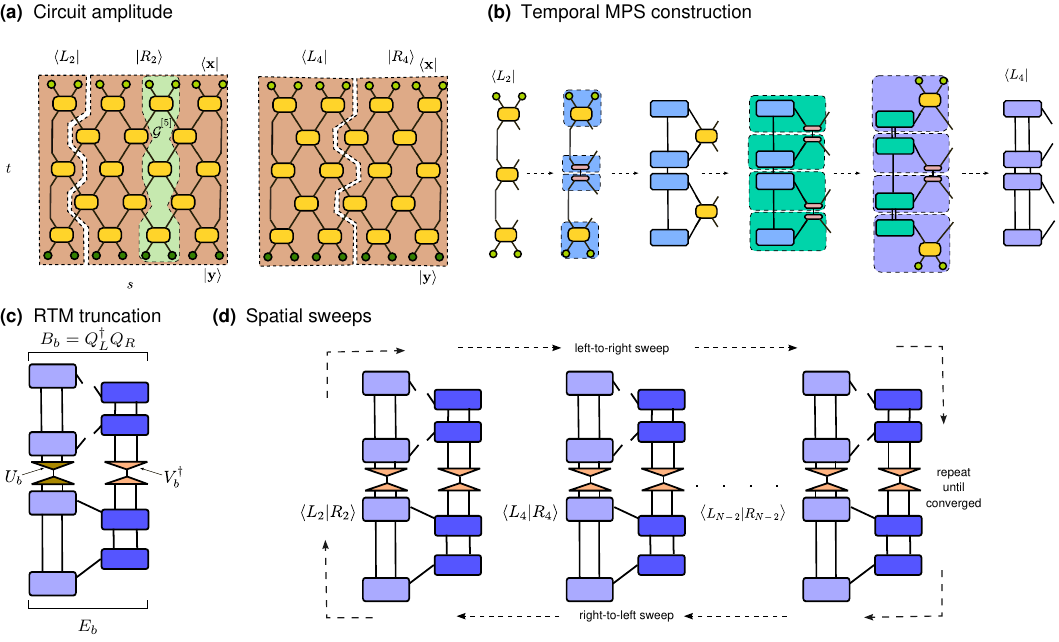}
 \caption{Sweeping RTM contraction.
 (a) The circuit amplitude between product states (green dots) is the
 overlap of two temporal boundaries at any spatial cut. Pale-green
 columns denote the transverse gate updates.
 (b) Absorbing further columns enlarges the temporal bonds.
 (c) Joint compression uses \rev{the singular vectors $U_b,V_b$ of $E_b$},
 selected by their contributions to the overlap, Eq.~\eqref{eq:overlap-weighted-score}.
 %\rev{The lower contraction is $B_b=Q_L^\dagger Q_R$.}
 (d) Repeated spatial sweeps update both boundaries and their environments;
 the implementation alternates center-out and boundary-in passes.
 Agreement across cuts, sweeps, and bond dimensions determines acceptance.}
 \label{fig:rtm-schematic}
\end{figure*}

\paragraph{Overlap-based compression.}
At a given spatial cut $j$,
we denote as $\bra{L_j}$ and $\ket{R_j}$ the temporal states obtained from the contraction of network to the left and the right of $j$, respectively.
Their exact %partial
contractions satisfy
$\A=\braket{L_j}{R_j}$ independently of $j$ [Fig.~\ref{fig:rtm-schematic}(a)].
The corresponding transition matrix built from them and its reduction across a temporal
bond $b$ are
\begin{equation}
 \T_j=\frac{\ket{R_j}\bra{L_j}}{\braket{L_j}{R_j}},\qquad
 \tau_b=\operatorname{Tr}_{\bar A_b}\T_j.
 \label{eq:transition-matrix}
\end{equation}
% $\tau_b$ is generally non-Hermitian.
We approximate both boundaries by tMPS with maximum bond dimension
$\chi$, applying the constituent transverse two-site gates directly,
without assembling a column MPO. After an update enlarges a bond,
we compress the two tMPS jointly [Fig.~\ref{fig:rtm-schematic}(b,c)].

$\tau_b$ is not hermitian in general, so our truncation is based on its singular value decomposition.
The crucial truncation criterion however is the contribution to the scalar
overlap, not the singular values alone. 
We begin by canonicalizing the tensors of the left and right vectors on
$A_b$ to isometries $Q_L,Q_R$.
The partial trace on $\bar A_b$  gives a $\chi \times \chi$ 
matrix $E_b$,
and
one has
$\tau_b=Q_RE_bQ_L^\dagger/\A_j$. Defining $B_b=Q_L^\dagger Q_R$ and decomposing $E_b=U_b\Sigma_bV_b^\dagger$ we have
\begin{equation}
 \A_j=\operatorname{Tr}(B_bE_b)=\sum_i \lambda_i,\qquad
 \lambda_i=\sigma_i v_i^\dagger B_bu_i.
 \label{eq:overlap-weighted-score}
\end{equation}
We retain at most $\chi$ singular directions with the largest
$|\lambda_i|$, then absorb their left and right singular vectors into the
respective boundaries. This retains directions important to the desired
overlap, including its environment-dependent cancellations.
The singular spectra and associated SVD entropy are examined in End Matter.

\begin{figure*}[t]
 \centering
 \includegraphics[width=\textwidth]{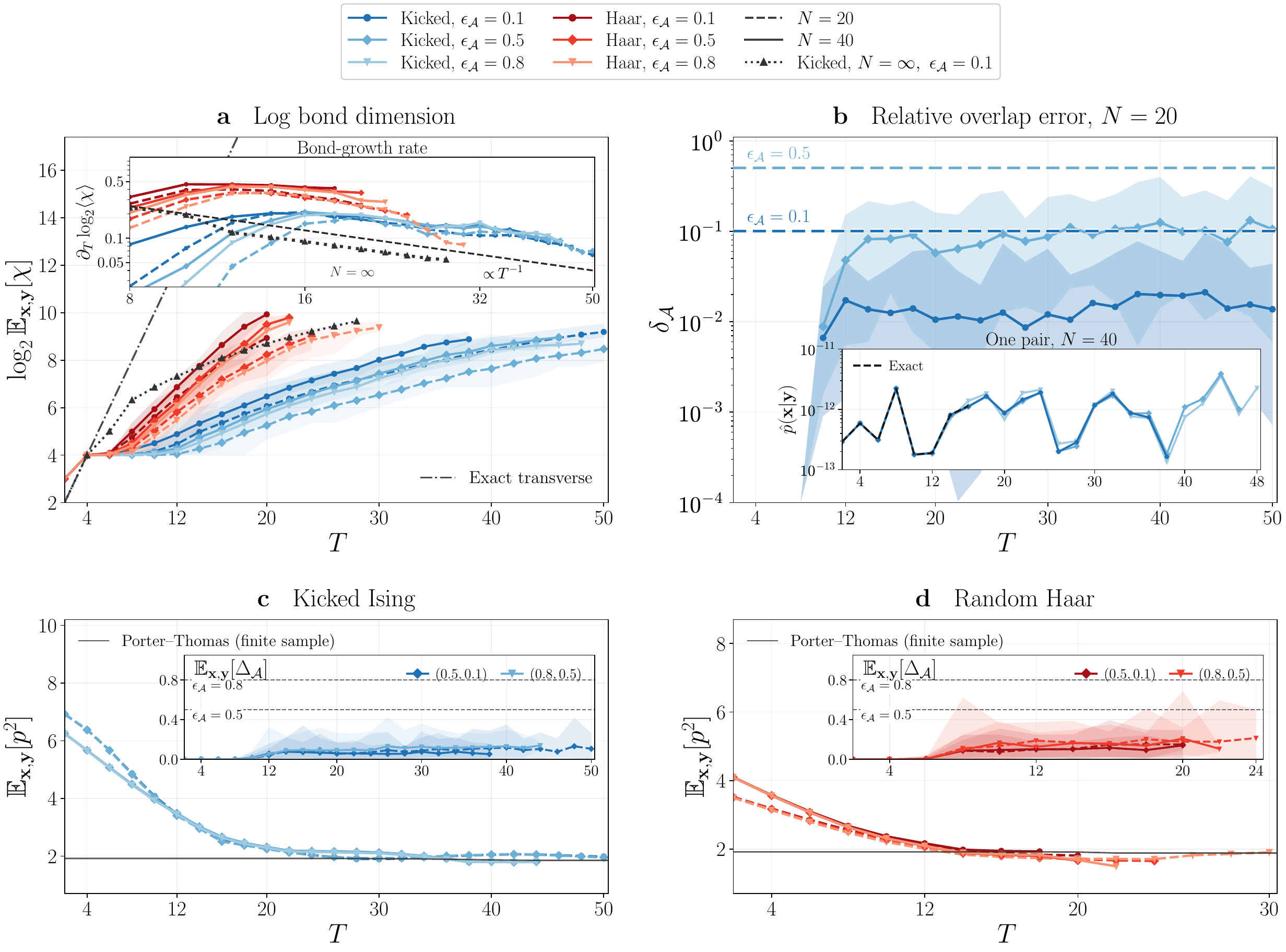}
 \caption{Strong-simulation diagnostics; \rev{in panels~(a,c,d), dashed and solid
 data curves denote $N=20$ and $N=40$.}
 (a) Base-two logarithm of the mean maximum retained bond dimension;
 shading spans the full pair range. \rev{The untruncated dimension-counting
 reference is $\chi=2^T$. The inset shows the numerical growth rate
 of $\log_2\E[\chi]$, matching pairs between adjacent times.} Gray triangles show the kicked-Ising
 $N=\infty$, $\mathbf0\to\mathbf0$ power-method calculation.
 Its inset rate is multiplied by $1/2$ for visibility.
 (b) True complex-amplitude error for kicked Ising at $N=20$
(see Fig.~\ref{fig:haar-exact-checks} for Haar circuits);
shading spans all errors, and dashed lines mark the targets.
The inset shows $\hat p(\bx|\by)$ for a fixed kicked-Ising pair
at $N=40$ at
$\epsilon_{\mathcal A}=0.1,0.5,0.8$, together with its exact value up to the largest computable time. 
 (c,d) Sample-mean-normalized second probability moment
 $\mathcal I_2$ [Eq.~\eqref{eq:sample-second-moment}], labeled $\E[p^2]$ in the plots,
 compared with its finite-sample Porter--Thomas mean.
 Insets compare amplitudes at targets $(\epsilon_{\mathcal A},\epsilon'_{\mathcal A})$
 on identical strings: $\Delta_{\mathcal A}=|\hat\A_{\epsilon_{\mathcal A}}-
 \hat\A_{\epsilon'_{\mathcal A}}|/|\hat\A_{\epsilon'_{\mathcal A}}|$.
 Shading gives full pair ranges; the guides are $0.5$ and $0.8$.}
 \label{fig:main-scaling}
\end{figure*}

\paragraph{Sweeps and error control.}
At fixed trial bond dimension, spatial sweeps across sites repeatedly update both left and right
boundaries. Exact cut amplitudes would coincide; their discrepancies
therefore test the approximate contraction due to truncation. For two complex numbers, define
$d(a,b)=|a-b|/\min(|a|,|b|)$, with $d(0,0)=0$ and
$d(a,0)=\infty$ when $a\ne0$. We monitor
\begin{equation}
 \begin{aligned}
 \Delta_{\rm cut}&=\max_{j,k}d(\A_j,\A_k),\\
 \Delta_{\rm sweep}&=\max_{j,k}d(\A_j,\A_k^{\rm prev}),\\
\Delta_\chi&=\max_{j,k}d(\A^\chi_j,\A_k^{\chi'}),
 \end{aligned}
 \label{eq:internal-errors}
\end{equation}
over different spatial cuts, sweeps and values of bond dimensions, demanding that they are below the thresold  $\epsilon_\mathcal{A}$.  We increase the trial maximum bond dimension along the empirical grid $\chi_n=16+n^2$ until consecutive
sweeps pass. The candidate is then frozen and validated for $N/2$ sweeps
at each of $\chi_n$ and $\chi_{n+1}$, requiring persistent agreement with
the candidate. Failure to stay below the thresold discard the candidate and resumes the search with a larger bond dimension. More details on the algorithm
are given in End Matter.

\rev{The circuit simulations use the complex-amplitude target
$\epsilon_{\mathcal A}$. For nonzero amplitudes, define the true relative
errors $\delta_{\mathcal A}=|\hat\A-\A|/|\A|$ and
$\delta_p=|\hat p-p|/p$, with $\hat p=|\hat\A|^2$.
The same internal tests can instead target probabilities by applying
$d$ to $|a|^2,|b|^2$ with threshold $\epsilon_p$.} If the true amplitude error satisfies
$\delta_{\mathcal A}\leq\epsilon_{\mathcal A}$, then
$\delta_p\leq2\epsilon_{\mathcal A}+\epsilon_{\mathcal A}^2$.
We stress that the sweep checks themselves do not prove such a bound or guarantee
convergence. Their reliability is a numerical finding that is tested below via different checks. 

\paragraph{Numerical results.}\leavevmode
\rev{We now report the result of strong simulations, see Fig.~\ref{fig:main-scaling}, of different chaotic open chains of
$N=20,40$ qubits}, with time $T$ counting
complete odd--even brick-wall periods. The chaotic kicked-Ising-type circuit
\cite{lerose2021} repeats the two-qubit gate
\begin{equation}
 \begin{aligned}
 U_{\rm KI}&=e^{-iH_X}e^{-iH_Z},\\
 H_X&=X\otimes X+0.81(X\otimes I+I\otimes X)/2,\\
 H_Z&=0.904508(Z\otimes I+I\otimes Z)/2,
 \end{aligned}
 \label{eq:kicked-ising-gate}
\end{equation}
where $X,Z$ are Pauli matrices. The random circuit uses independent
Haar gates in $U(4)$ at every bond, layer, and period. We uniformly sample up to 24
computational-basis input--output pairs with $\by\ne\bx$;
each Haar pair has its own circuit realization.

Figure~\ref{fig:main-scaling}(a) shows $h_\chi(T)=\log_2\E[\chi(T)]$. Fixed exponential growth would give a straight
line and a constant positive derivative. Instead, the curves bend downward
and their growth rates decrease, consistent with subexponential growth
over the measured window. \rev{The late-time growth rate is compatible with
$T^{-1}$, with a crossover that depends on the target
$\epsilon_{\mathcal A}$. The translationally invariant, infinite
kicked-Ising chain with $\bx=\by=\mathbf0$, accessible by a power method
for the transverse transfer operator, displays a clearer approximately
logarithmic increase in $h_\chi(T)$ over the displayed interval.}
Its RTM entropy shows a similar trend, recalling continuous-time results
\cite{carignano2025cft,carignano2025sampled}; finite-size generic-string
contractions also probe nonstationary transfer modes.

Panel~(b) compares the approximate complex amplitudes with exact
forward evolution for kicked Ising at $N=20$. Every completed pair remains
below its target $\epsilon_{\mathcal A}=0.1$ or $0.5$ over the displayed
window. This is stronger than agreement of the mean alone, since the
shaded envelope includes the full range. The insets of panels~(c,d)
compare amplitudes at different targets on identical strings, providing
an additional consistency check at both sizes.

The statistics of the sampled probabilities provide \rev{a complementary test}. For $M$
uniformly queried pairs, rather than Born-distributed samples, we form
\begin{equation}
 \mathcal I_2=M\frac{\sum_{i=1}^M\hat p_i^2}
                         {(\sum_{i=1}^M\hat p_i)^2},\qquad
 \mathbb E_{\rm PT}\mathcal I_2=\frac{2M}{M+1}.
 \label{eq:sample-second-moment}
\end{equation}
The finite-sample reference follows because independent exponential
weights, normalized by their sum, have a Dirichlet distribution with unit
parameters \cite{porter1956,boixo2018}. The moment approaches 2 as
$M\to\infty$. For both circuits, the late-time data approach the
Porter--Thomas scale, with finite-sample fluctuations and residual
circuit-dependent deviations. \rev{Together with the cross-target
comparisons, this agreement supports the stability of the computed
probability statistics deep in the anticoncentration regime.}

\begin{figure}[!t]
 \centering
 {\raggedright\textbf{(a)}\par}
 \includegraphics[width=\linewidth]{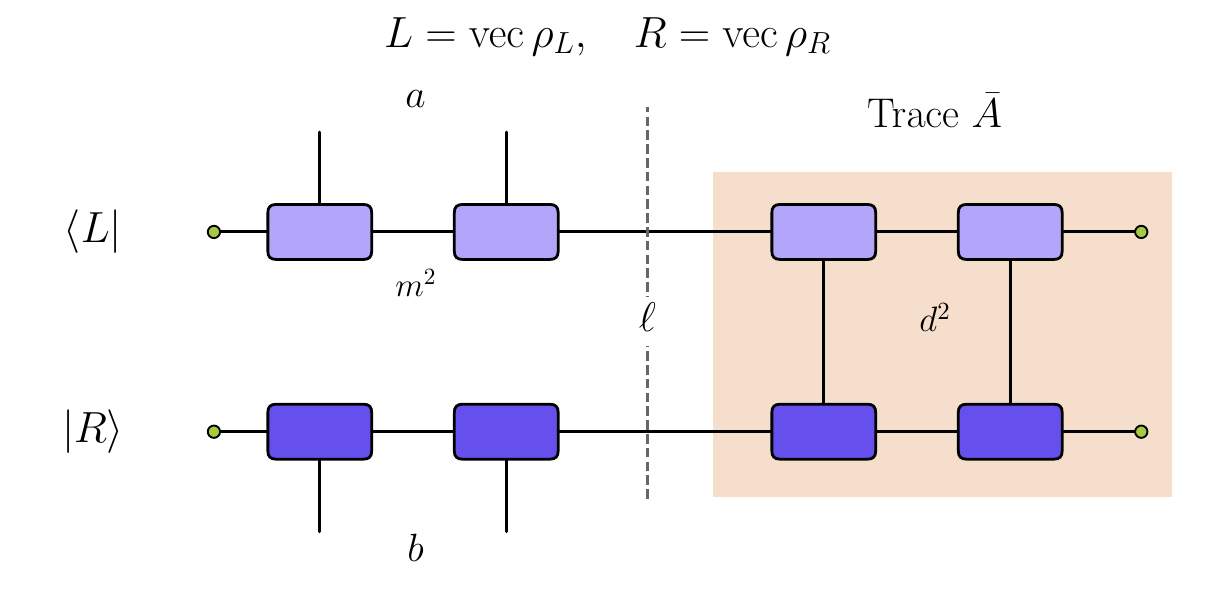}\par
 \smallskip
 {\raggedright\textbf{(b)}\par}
 \includegraphics[width=0.90\linewidth]{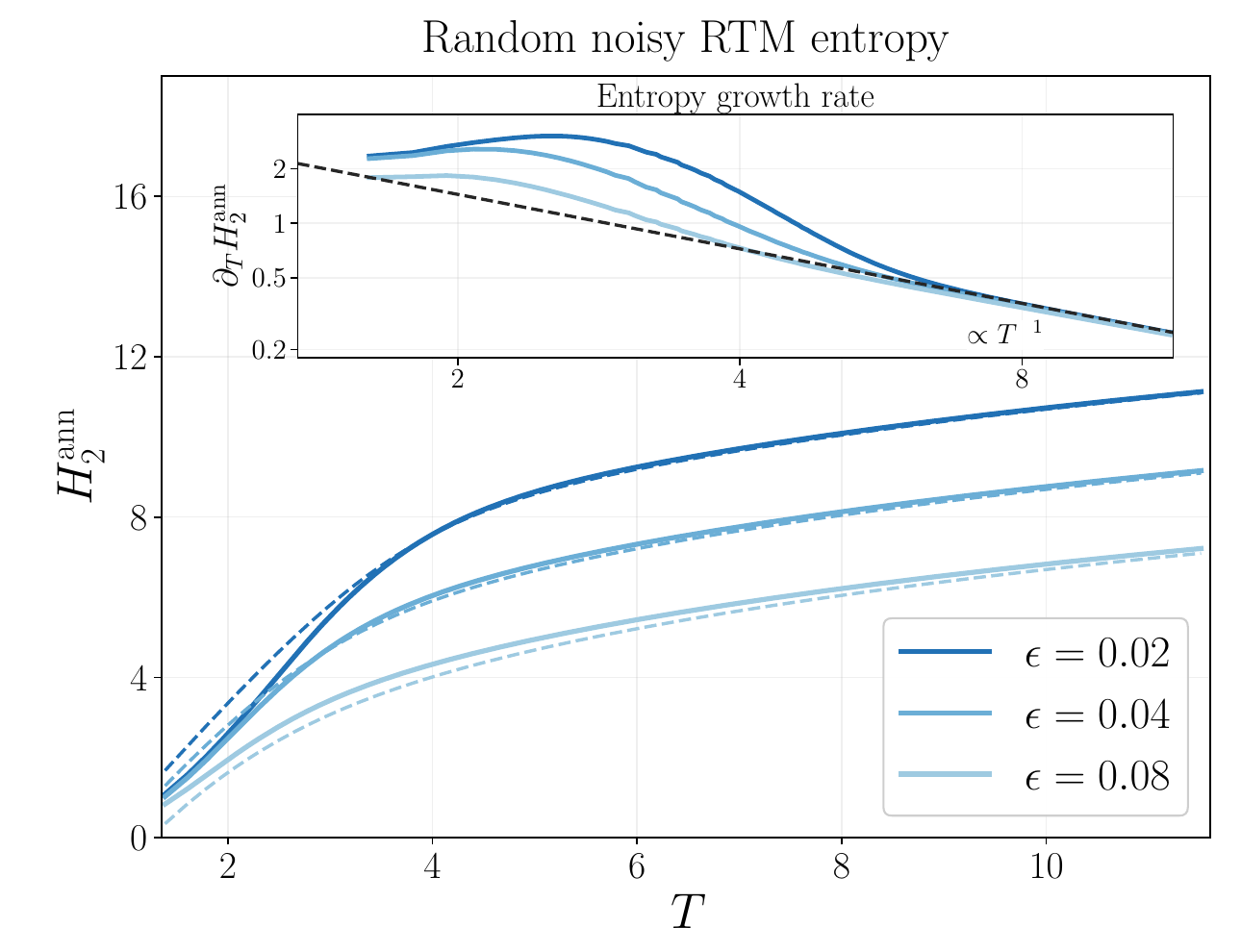}
 \caption{Noisy RTM proxy.
 (a) Two independent vectorized MPOs define the central RTM by tracing
 the shaded region, with $\ell=n/2$; green dots denote the memory boundaries.
 (b) Stationary annealed entropy for $d=2$, $T=\ln m$, $\gamma=\epsilon/T$.
 Solid curves are full replica-transfer results; dashed curves show
 the leading crossover approximation. The inset shows the growth rate
 and the asymptote $2/(T\ln2)$.}
 \label{fig:annealed-rtm}
\end{figure}

\paragraph{A random noisy RTM toy-model.}
To understand how a fixed accuracy requirement can curb temporal
complexity, \rev{we construct an RTM from two independent noisy random
matrix-product operators (RMPOs)}, following Ref.~\cite{sauliere2026anticoncentration}.
Each is generated by a Haar staircase with local dimension $d$, auxiliary
memory dimension $m\sim e^{T/\tau}$, and depolarizing strength $\gamma$.
The staircase length is first taken to stationarity at fixed $m,\gamma$,
and $m$ is then increased. Vectorizing the two operators gives proxy
boundaries $\ket{L},\ket{R}$ in a doubled Hilbert space; \rev{tracing out
half of the sites} defines the central RTM [Fig.~\ref{fig:annealed-rtm}(a)].

For their unnormalized central RTM $\tau_c$, define
$Z_k=\operatorname{Tr}[(\tau_c\tau_c^\dagger)^k]$ and
\begin{equation}
 \mathcal P_{\rm ann}=\frac{\mathbb E_U Z_2}{\mathbb E_U(Z_1^2)},\qquad
 H_2^{\rm ann}=-\log \mathcal P_{\rm ann}.
 \label{eq:toy-annealed}
\end{equation}
 Haar averaging
reduces its moments to a replica-transfer problem. In the weak-noise,
large-memory regime $m\gamma\gg1$, the calculation in End Matter gives
$\mathcal P_{
\rm ann}\sim C_d\gamma^2$, where $C_d$ depends only on $d$.
\rev{To model fixed relative precision phenomenologically, we set
$\gamma=\epsilon/T$, keeping the accumulated noise scale
$\gamma T=\epsilon$ fixed. Here $\epsilon$ is a proxy-model parameter,
not the algorithmic tolerance $\epsilon_{\mathcal A}$.} This yields
\begin{equation}
 H_2^{\rm ann}\sim \log (1/\gamma)\sim\log(T/\epsilon).
 \label{eq:toy-stationary}
\end{equation}
Initially, the entropy grows approximately linearly; the crossover to logarithmic growth occurs when $m\gamma$ becomes of order one,
or $\epsilon e^{T/\tau}/T\sim1$. Figure~\ref{fig:annealed-rtm}(b) displays
this crossover and the late-time growth rate $2/(T\ln2)$. This phenomenological model shows how exponentially
growing memory can coexist with only logarithmic RTM entropy at finite relative
precision. \rev{In the circuit contraction, the exact state's entanglement is not reduced:}
only information unnecessary for the targeted overlap is discarded.

\paragraph{Applications and outlook.}\leavevmode
\rev{A direct application of our findindigs is that benchmarking large-scale quantum computations, at least for one-dimensional-like geometries, is possible:
relative-accuracy queries enable linear cross-entropy benchmarking}
(XEB) without storing the full ideal state \cite{boixo2018,arute2019}.
For $M$ measured device strings, let
$X=2^NM^{-1}\sum_i p(\bx_i|\by)-1$, and let $\hat X$ replace
$p$ by the RTM estimates on the same strings. \rev{If each queried probability
satisfies $\delta_p\leq\epsilon_p$,}
\begin{equation}
 |\hat X-X|\leq\epsilon_p(X+1).
 \label{eq:xeb-error}
\end{equation}
Where XEB tracks the device fidelity $\mathcal F$, this gives
classical corrections $\pm\epsilon_p(\mathcal F+1)$, in addition to
measurement uncertainty \cite{ware2023xeb,sauliere2026anticoncentration}.
Complex-amplitude queries also supply the conditional states used in
shadow-overlap certification \cite{huang2024certifying}. For
likelihood-based learning \cite{liu2018born}, a true bound
$\delta_p\leq\epsilon_p<1$ controls each base-two log-probability error
by $-\log_2(1-\epsilon_p)$, even for exponentially small probabilities.

Our central result is strong numerical evidence that finite relative
precision can make individual probability queries substantially cheaper
than a faithful state representation. The main open question, left for future work, is whether the internal checks can be promoted to
rigorous and controlled error bounds. The noisy RTM model provides a
possible mechanism for the logarithmic growth of the RTM entropy; however, a more direct relation to the actual sweep dynamics requires
further work. Variational overlap optimization
\cite{haegeman2011,SciPostPhysCore.4.1.004} may also be a more direct implementation route. Finally,
an extension to two spatial dimensions is clearly the next step and is currently under development \cite{lubasch2014,schuch2007}.
\label{letter-main-end}

\begin{acknowledgments}
\paragraph{Acknowledgments.}
We thank Guglielmo Lami and Arman Sauliere for valuable input and collaboration on
related projects, and Mari Carmen Ba\~nuls, Bruno Bertini, and Tianci Zhou
for discussions.
J.D.N. and M.G. acknowledge support from the European Research Council (ERC)
Starting Grant No.~101042293 (HEPIQ) and the Agence Nationale de la
Recherche (ANR) under grant ANR-22-CPJ1-0021-01.
L.T. acknowledges ongoing discussions on related topics with the
group at CSIC, including Aleix Bou Comas, Francois Payn, Jan Thorben
Schneider, Esperanza Lopez, and Sergio Cerezo Roquebrun, and support from
the Proyecto Sin\'ergico CAM Programa
TEC-2024/COM-84 QUITEMAD-CM; the CSIC Research Platform on Quantum
Technologies PTI-001; and grant PID2024-160172NB-I00, funded by
MICIU/AEI/10.13039/501100011033 and FEDER/UE.
Computing resources were provided in part by the Centre de Calcul de
CY Cergy Paris Universit\'e.
S.C. acknowledges support from his AI4S fellowship within the
``Generaci\'on D'' talent-attraction initiative of Red.es,
Ministerio para la Transformaci\'on Digital y de la Funci\'on P\'ublica
(C005/24-ED CV1), funded by NextGenerationEU through PRTR.
We also acknowledge support from grant PID2023-151384NB-I00, financed by
MCIU/AEI/10.13039/501100011033 and FSE+.
\end{acknowledgments}

\bibliographystyle{apsrev4-2}
\bibliography{references}

\onecolumngrid
\clearpage
\begin{center}\textbf{END MATTER}\end{center}
\label{letter-end-matter-start}
\begin{figure}[h!]
 \centering
 \includegraphics[width=\textwidth]{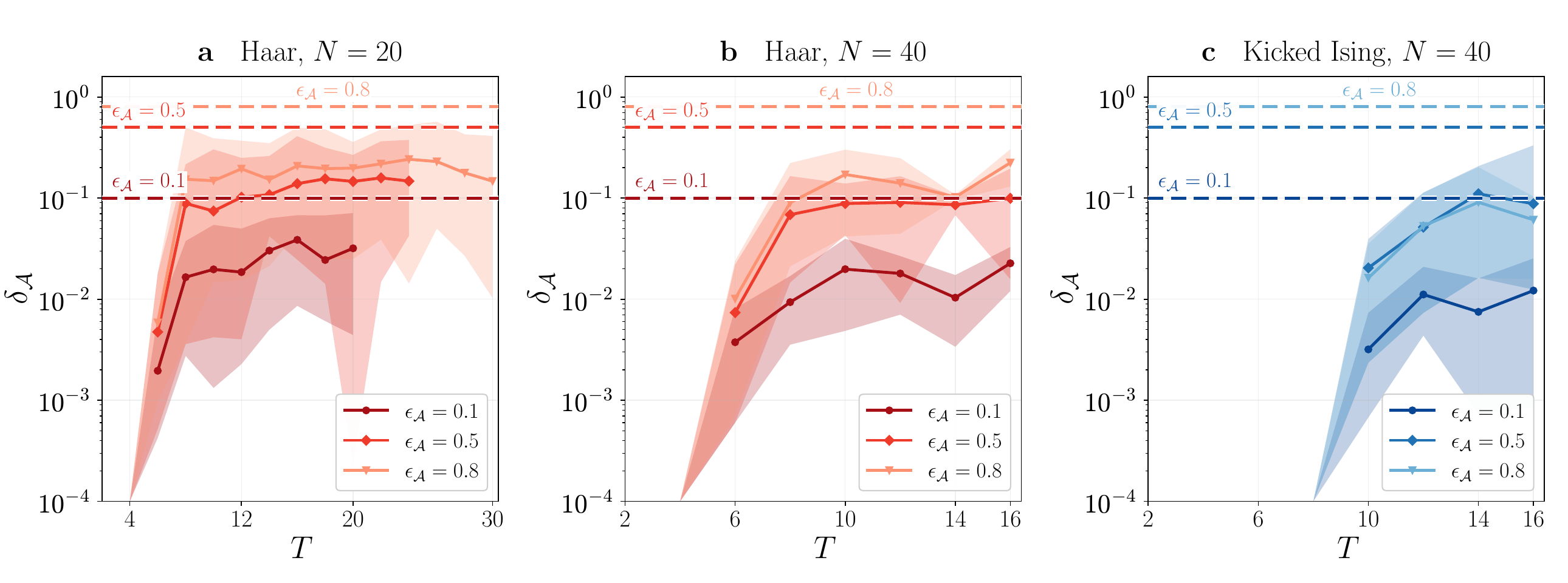}
 \caption{Additional exact-reference checks for independent Haar circuits
at (a) $N=20$ and (b) $N=40$, and for the kicked-Ising circuit at
(c) $N=40$. The relative complex-amplitude error is
$\delta_{\mathcal A}=|\hat{\mathcal A}-\mathcal A_{\rm exact}|/
|\mathcal A_{\rm exact}|$. Solid curves show the arithmetic mean
over the tested input--output pairs, and shading spans the full
minimum--maximum range, without time smoothing. Dashed horizontal
lines mark the targets $\epsilon_{\mathcal A}=0.1,0.5,0.8$.
Exact references are obtained by full-statevector evolution in (a)
and untruncated scalar tensor-network contraction in (b,c), reaching
$T=16$ for both $N=40$ circuits. All tested errors remain below their
respective targets. Errors below $10^{-4}$ lie outside the displayed
vertical range.}
 \label{fig:haar-exact-checks}
\end{figure}
\twocolumngrid

\paragraph{{Additional numerical checks.}}
\rev{Figure~\ref{fig:haar-exact-checks} compares the Haar-circuit RTM
amplitudes in Fig.~\ref{fig:main-scaling} with independent exact
references. At $N=20$, the references are obtained by evolving the full
$2^{20}$-component statevector, without truncation. At $N=40$, we instead
contract the tensor network for the specified complex amplitude directly,
fixing both computational-basis boundaries and optimizing the binary
contraction order with \texttt{cotengra}~\cite{gray2021a}.
}

\begin{figure*}[t]
 \centering
 \includegraphics[width=0.94\textwidth]{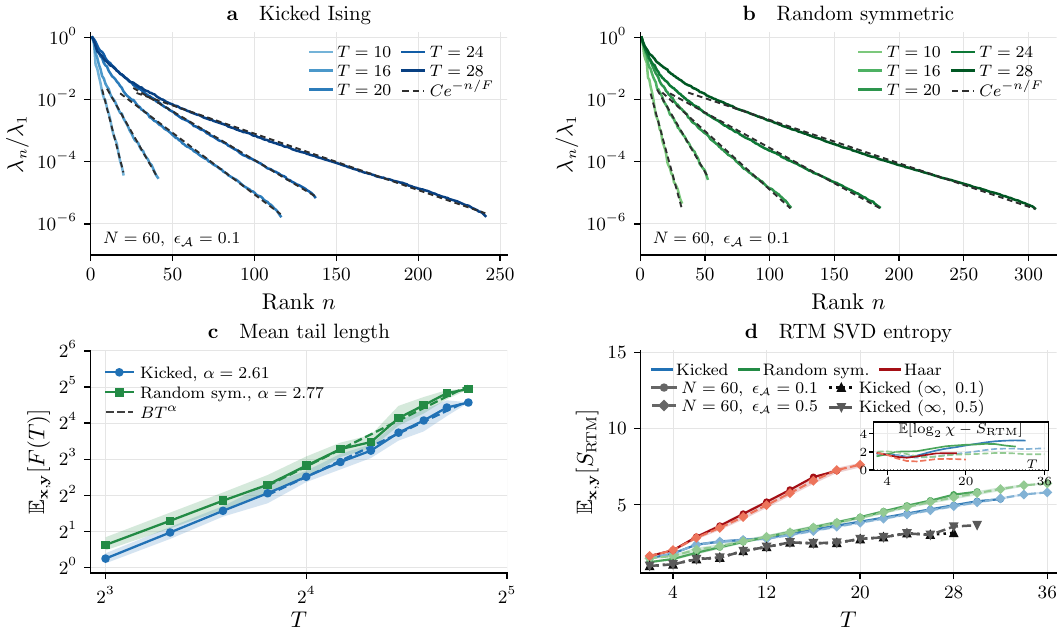}
 \caption{\textbf{RTM spectra and entropy}.
 (a,b) Single-cut spectra $\sigma_n/\sigma_1$ for kicked Ising and random
 symmetric circuits at $N=60$, $\epsilon_{\mathcal A}=0.1$, at different times.
 Darker shades indicate later times; dashed segments fit the resolved
 exponential tails. (c) Mean decay length in both models, averaged over
 bitstrings, with power-law fits. (d) Mean maximum local RTM SVD entropy
 for the three $N=60$ circuit ensembles at $\epsilon_{\mathcal A}=0.1,0.5$.
 Black/gray triangles show translationally invariant kicked-Ising
 $N=\infty$, $\bx=\by=\mathbf0$ results at $\epsilon_{\mathcal A}=0.1,0.5$.
 The inset shows the gap to the entropy bound,
 $\E[\log_2\chi-S_{\rm RTM}]$, for the finite-size data,
 with the same colors and line styles.}
 \label{fig:singular-mechanism}
\end{figure*}

\paragraph{RTM spectra and entropy.}
Here we present additional numerical results on the singular-value spectrum.
Denoting the singular values of $\tau_b$ by $\{\sigma_n\}$, we define
the RTM SVD entropy using base-two logarithms,
\begin{equation}
 S_{\rm RTM}=-\sum_n w_n\log_2 w_n,\qquad
 w_n=\frac{\sigma_n}{\sum_m\sigma_m}.
 \label{eq:entropy-from-spectrum}
\end{equation}
This quantity, also studied in Ref.~\cite{li2026rtm}, is distinct from the
temporal entanglement entropy of either boundary separately
\cite{parzygnat2023,carignano2024rtm}.
The $\sigma_n$ are nonnegative singular values, not the complex
overlap contributions $\lambda_i$ in Eq.~\eqref{eq:overlap-weighted-score}.
For canonical $Q_L,Q_R$, the singular values of $E_b$ differ from those of
$\tau_b$ by the common factor $|\A_j|$, which cancels in the normalized
weights and spectral ratios.
\rev{For the additional exchange-symmetric random ensemble in
Fig.~\ref{fig:singular-mechanism}, we draw $U_0$ from Haar measure on
$U(4)$ and form $\widetilde U=(U_0+P_{\rm sw}U_0P_{\rm sw})/2$,
where $P_{\rm sw}$ swaps the two qubits. Its polar unitary is
$U_{\rm sym}=VW^\dagger$, with
$\widetilde U=V\Sigma W^\dagger$, so that
$[U_{\rm sym},P_{\rm sw}]=0$. Gates are drawn independently at each
bond, layer, and period.}

Over the accessible time window, the resolved tails are approximately
described by
\begin{equation}
 \frac{\sigma_n(T)}{\sigma_1(T)}\simeq C(T)e^{-n/F(T)}.
 \label{eq:exponential-tail}
\end{equation}
Here $F(T)$ is the characteristic decay scale, extracted from a linear fit
to the logarithm of the tail. Each displayed spectrum uses one bit-string
pair and one central spatial and temporal cut. Panel~(c) summarizes the
decay scale averaged over the sampled input--output pairs. Over the
accessible time window, its mean is approximately described by
$\E[F(T)]\simeq BT^\alpha$.
If the full normalized spectrum were geometric,
$w_n=(1-q)q^{n-1}$ with $q=e^{-1/F}$, then for $F\gg1$,
$S_{\rm RTM}=\log_2F+O(1)$. Thus $F\propto T^\alpha$ would imply
$S_{\rm RTM}=\alpha\log_2T+O(1)$.
The spectra in panels~(a,b), however, have additional structure among
their largest singular values, outside the fitted tail. The evolving
weight of these values can produce deviations from the logarithmic
behavior suggested by the fitted tail alone.
On the other hand, the infinite, translationally invariant $\mathbf0\to\mathbf0$
calculation is more consistent with pure logarithmic growth.
For a retained RTM spectrum of rank $\chi$, $S_{\rm RTM}\leq\log_2\chi$.
\rev{Panel~(d) averages each run's maximum local entropy. The inset
uses the same runs to average $g=\log_2\chi-S_{\rm RTM}$, with $\chi$
the maximum retained bond dimension. This is the mean of the per-run
gap, not $\log_2\E[\chi]-\E[S_{\rm RTM}]$.
Both curves use a three-neighbor time average. Late-time mean gaps
remain of order one to three bits, corresponding to a geometric-mean
ratio $\chi/2^{S_{\rm RTM}}$ of order two to ten.}

\rev{The two exact-reference implementations agree to better than
$3\times10^{-13}$ relative error in an independent $N=20$ test through
$T=12$. All plotted RTM errors lie below their respective targets.
The $N=20$ comparison includes 24 independent circuit/pair realizations
per available time and target. At $N=40$, ten realizations are checked
through $T=12$, with additional checks of two of them at $T=14$;
the same realizations are used for all three targets.
The shorter $N=40$ reference window reflects the cost of exact
contraction, not the endpoint of the RTM data.}

\paragraph{More details on the numerical procedure and local truncation.}
For fixed input and output bit strings $\by$ and $\bx$, we first absorb
the physical boundary indices into the circuit gates. The resulting
space-time tensor network can then be viewed as a contraction of spatial
columns of tensors. We denote the column at site $j$ by
$\mathcal G^{[j]}_{x_jy_j}$, omitting the configuration indices below
for brevity. Representative columns are shaded in pale green in
Fig.~\ref{fig:rtm-schematic}(a). \rev{For two spatial cuts $j<k$, the
amplitude can be written schematically as}
\begin{equation}
 \rev{\A_{\bx\by}(T)=\bra{L_j}
 \mathcal G^{[j+1]}\cdots\mathcal G^{[k]}\ket{R_k},}
 \label{eq:transverse-product}
\end{equation}
\rev{where the boundary vectors contain the contractions outside the
interval between these cuts.} Equation~\eqref{eq:transverse-product} is a compact
notation: in the implementation we do not explicitly construct a column
as an MPO, but apply its constituent transverse two-site gates directly.
At every spatial cut the exact overlap $\braket{L_j}{R_j}$ is the same.
Constructing these boundaries exactly is exponentially costly, however,
and truncating them introduces discrepancies between cuts.

The RTM truncation proceeds as follows. Consider a temporal bond $b$ of
the current pair $\bra{L_j},\ket{R_j}$, and split the temporal chain into
$A_b$ and $\bar A_b$. Separately canonicalizing the tensors on $A_b$
gives isometries satisfying $Q_L^\dagger Q_L=Q_R^\dagger Q_R=I$.
Contracting $\bar A_b$ leaves the two-index overlap matrix $E_b$, with
\begin{equation}
 \tau_b=\frac{Q_RE_bQ_L^\dagger}{\A_j},\qquad
 B_b=Q_L^\dagger Q_R,\qquad
 \A_j=\operatorname{Tr}(B_bE_b).
 \label{eq:local-rtm}
\end{equation}
We first compute the full SVD $E_b=U_b\Sigma_bV_b^\dagger$.
Crucially, the truncation does not simply keep the largest singular values.
Each direction contributes the complex number $\lambda_i$ defined in
Eq.~\eqref{eq:overlap-weighted-score}. For a given maximum bond dimension
$\chi$, we choose the set $\mathcal K_\chi$ corresponding to the
$\min(\chi,\dim\Sigma_b)$ largest scores $|\lambda_i|$ and replace
\begin{equation}
 \begin{aligned}
 E_b&\longmapsto E_b^{(\chi)}
       =\sum_{i\in\mathcal K_\chi}\sigma_i u_iv_i^\dagger,\\
 \A_{j,\mathrm{kept}}&=\sum_{i\in\mathcal K_\chi}\lambda_i.
 \end{aligned}
 \label{eq:weighted-projection}
\end{equation}
Here $\sigma_i$ denotes a singular value of the unnormalized matrix $E_b$,
as in Eq.~\eqref{eq:overlap-weighted-score}. The selected columns of
both $U_b$ and $V_b$ are absorbed into the two temporal boundaries and
their neighboring tensors [Fig.~\ref{fig:rtm-schematic}(c)]. For a fixed
bond dimension, we repeat the local updates in alternating center-out
and boundary-in spatial sweeps, implementing the repeated optimization
sketched in Fig.~\ref{fig:rtm-schematic}(d).

\paragraph{Search and validation schedule.}
After every update, we record the overlap
\begin{equation}
 \A_j^{(\chi)}=\braket{L_j^{(\chi)}}{R_j^{(\chi)}}
 \label{eq:cut-amplitudes}
\end{equation}
at the visited spatial cut $j$. A completed sweep provides values at
every recorded cut. Their spread provides a first internal consistency
check of the approximate contraction. The diagnostics in
Eq.~\eqref{eq:internal-errors} compare all recorded cuts, not just
matching indices: $\Delta_{\rm cut}$ measures agreement within a sweep,
$\Delta_{\rm sweep}$ compares consecutive sweeps, and $\Delta_\chi$
compares the validation amplitudes with the frozen candidate.
A test passes when its relative distance is below the prescribed
target $\epsilon_{\mathcal A}$. Probability targeting instead applies
the same distance to $|a|^2$ and $|b|^2$, with target $\epsilon_p$.

In this work, we use the trial bond-dimension grid $\chi_n=16+n^2$.
At each grid value, we perform at least four search sweeps and require
two consecutive sweeps for which both $\Delta_{\rm cut}$ and
$\Delta_{\rm sweep}$ pass. If at least one of the first four sweeps
passes, the search may continue to ten sweeps at that cap, stopping
once the consecutive-pass condition is met. Otherwise the cap is increased.
A candidate is then frozen and subjected to $N/2$ all-passing validation
sweeps at $\chi_n$, followed by $N/2$ at $\chi_{n+1}$. Each validation
sweep is compared with the same frozen candidate; we also check pairwise
agreement across the validation cuts and sweeps. A failure rejects the
candidate and resumes the search with the evolved environments.

After successful validation, we report the original candidate,
\begin{equation}
 \hat p(\bx|\by)=|\A^{\rm frozen}_{N/2}|^2,
 \label{eq:accepted-probability}
\end{equation}
and its maximum actual retained bond dimension $\chi\leq\chi_n$,
excluding the larger cap visited only for validation. Subsequent times
inherit $\chi_n$ and the accepted environments; two random tensors are
inserted per added circuit period to extend each temporal chain. Thus
the reported resource curves include this sequential tensor warm start.
The internal checks do not provide rigorous error bounds or a proof of
convergence; their reliability is assessed against exact amplitudes and
through the cross-target comparisons.

\paragraph{More details on the noisy RTM construction.}
We consider two independent noisy Haar staircases, following the RMPO
construction of Ref.~\cite{sauliere2026anticoncentration}. Each staircase
contains $n\gg1$ emitted sites of fixed local dimension $d$ and an auxiliary
bond of dimension $m\sim e^{T/\tau}$ that acts as a temporal memory.
Here $\tau$ sets the effective timescale for the growth of the memory
dimension. To avoid nonuniversal boundary corrections, we first take
the large-$n$ stationary limit at fixed $m,\gamma$, and then let $m$
increase. At step $r$, a fresh site initialized in $\ket0\bra0$
interacts with the memory through an independent $U_r\in U(md)$,
followed by depolarization:
\begin{equation}
 \mathcal E_{\gamma,U_r}(X)=(1-\gamma)U_rXU_r^\dagger
       +\frac{\gamma}{md}\operatorname{Tr}(X)I_{md}.
 \label{eq:toy-channel}
\end{equation}
The memory is initialized in $\ket0$ and projected back onto this state
at the end. The resulting unnormalized density MPO $\rho$ has a local tensor
\begin{equation}
 \begin{aligned}
 \mathcal B^{st}_{\alpha\alpha',\beta\beta'}
 ={}&(1-\gamma)U_{s\alpha,0\beta}U^*_{t\alpha',0\beta'}\\
 &+\frac{\gamma}{md}
   \delta_{st}\delta_{\alpha\alpha'}\delta_{\beta\beta'}.
 \end{aligned}
 \label{eq:toy-tensor}
\end{equation}
Its physical index combines $(s,t)$ into dimension $d^2$, and its virtual
index combines two memory indices into dimension $m^2$. Two independent
draws define $\ket L=\operatorname{vec}\rho_L$ and
$\ket R=\operatorname{vec}\rho_R$. These are vectors in a doubled
Hilbert space, even though the underlying density operators are
generally mixed.

To compute the RTM at the central cut, we contract the last $n-\ell$
sites, with $\ell=n/2$. We denote this unnormalized central reduction by
\begin{equation}
 \tau_c=\operatorname{Tr}_{\bar A}\ket R\bra L.
 \label{eq:toy-central-rtm}
\end{equation}
The shaded region in Fig.~\ref{fig:annealed-rtm}(a) is contracted,
while the open indices on $A$ are retained. Define
$Z_k=\operatorname{Tr}[(\tau_c\tau_c^\dagger)^k]$ as in
Eq.~\eqref{eq:toy-annealed}. Here $\mathbb E_U$ denotes averaging over
both Haar staircases, without first normalizing each postselected MPO.
The annealed ratio probes weights proportional to squared singular values;
it is distinct from both the mean sample R\'enyi entropy and the
linear-singular-value entropy in Eq.~\eqref{eq:entropy-from-spectrum}.

\paragraph{Replica calculation of the annealed entropy.}
Both moments in Eq.~\eqref{eq:toy-annealed} can be computed by Haar
averaging and expressed through an effective one-dimensional transfer
matrix, following the framework of Ref.~\cite{sauliere2026anticoncentration}.
For four replicas, the normalized transfer of one staircase is, to
leading order in $m^{-1}$ and $\gamma$,
\begin{equation}
 \mathsf B\simeq\exp\!\left[\frac{d-1}{md}\mathsf A-\gamma Q\right].
 \label{eq:toy-replica-transfer}
\end{equation}
Here $\mathsf A$ is the adjacency matrix on permutations in $S_4$, with
$\mathsf A_{\pi\sigma}=1$ when $\pi$ and $\sigma$ differ by one
transposition, and $Q_\pi=4-\#\mathrm{Fix}(\pi)$ counts the replica
labels moved by $\pi$. On the shaded, traced side of the cut, the two
staircases are sewn at every site by
$C_{\pi\sigma}=d^{-|\pi^{-1}\sigma|}$, where $|\rho|$ is the
transposition length of $\rho$. On the unshaded side, each staircase
instead contracts separately with the twist imposed by the moment.

Let $e$ be the identity. The two cut twists can be drawn as
\begin{equation}
 \begin{aligned}
 a=(12)(34):&\quad
   \overbrace{1\,2}^{(12)}\quad\overbrace{3\,4}^{(34)},\\
 b=(14)(23):&\quad
   \overbrace{1\;\underbrace{2\,3}_{(23)}\;4}^{(14)}.
 \end{aligned}
 \label{eq:toy-twists}
\end{equation}
The denominator uses matched twists $(a,a)$, whereas the numerator
uses $(a,b)$. Following the transfer matrix from the cut into the traced
region, we obtain the following representative shortest permutation
histories for the numerator and denominator of Eq.~\eqref{eq:toy-annealed}:
\begin{equation}
 \begin{aligned}
 \mathrm{den}:&\quad (a,a)\xrightarrow{J}(t,t)\xrightarrow{J}(e,e),\\
 \mathrm{num}:&\quad (a,b)\to(a,u)\to(t,u)\\
 &\qquad\to(e,u)\to(e,e).
 \end{aligned}
 \label{eq:toy-domain-paths}
\end{equation}
In the denominator, $t=(12)$ or $(34)$. The numerator line shows one
ordering with $t=(12)$ and $u=(14)$; the other orderings enter the
coefficient below.

In the denominator, the matched $(a,a)$ and $(t,t)$ domains have
$C_{\pi\pi}=1$. Their lengths $r$ and $s$ are therefore cut off only
by noise, with weights $e^{-8\gamma r}$ and $e^{-4\gamma s}$.
Changing both staircases by the same transposition has weight
\begin{equation}
 J\simeq\left(\frac{d-1}{md}\right)^2
       \left(1+2\sum_{h\geq1}d^{-h}\right)=\frac{1-d^{-2}}{m^2}.
 \label{eq:toy-wall-weight}
\end{equation}
The sum also accounts for short intermediate domains in which the
staircases are mismatched. Writing $\widetilde D$ (denominator) and
$\widetilde N$ (numerator) for the two moments after removing their
common dimensional and stationary-transfer factors gives
\begin{equation}
 \begin{aligned}
 \widetilde D&\sim2J^2\sum_{r,s\geq0}e^{-8\gamma r-4\gamma s}\\
 &\sim\frac{(1-d^{-2})^2}{16m^4\gamma^2},
 \end{aligned}
 \label{eq:toy-den-walls}
\end{equation}
in the limit $\gamma\ll1$.
For the numerator, by contrast, the shortest routes from $a$ and $b$
to $e$ share only $e$. Every nonidentity pair on these routes is
therefore mismatched, and a sewing penalty is incurred at each site.
Along the ordering in Eq.~\eqref{eq:toy-domain-paths}, the four
nonidentity domains have sewing weights per site $d^{-2}$, $d^{-3}$,
$d^{-2}$, and $d^{-1}$, respectively. Their position sums remain finite
in the limit $\gamma\ll1$, even though they also carry noise weights.
Summing all shortest routes and the unshaded-side contractions yields
\begin{equation}
 \widetilde N\sim\frac{K_d}{m^4},\qquad
 K_d=(1+d^{-1})^4+d^{-2}.
 \label{eq:toy-num-walls}
\end{equation}
Thus, for small but finite $\gamma$, we obtain
$\mathcal P_{\rm ann}\sim C_d\gamma^2$, with
$C_d=16K_d/(1-d^{-2})^2$ and $C_2=1360/9$.
These asymptotics take $n\to\infty$ first, followed by large $m$
and weak noise with $m\gamma\gg1$.

We can now translate this result into the effective time dependence
by taking $m\sim e^{T/\tau}$ and $\gamma=\epsilon/T$. With the base-two
logarithm in Eq.~\eqref{eq:toy-annealed}, this gives
\begin{equation}
 H_2^{\rm ann}
 =2\log_2(T/\epsilon)-\log_2 C_d+o(1).
 \label{eq:toy-stationary-prefactor}
\end{equation}

\end{document}